\documentclass[10pt,journal,letterpaper]{IEEEtran}
\usepackage[T1]{fontenc}
\usepackage{amsmath,amssymb}
\usepackage[nocompress]{cite}
\usepackage{booktabs}
\usepackage{tabularx}
\usepackage{microtype}
\usepackage{xspace}
\usepackage{multirow}
\usepackage{graphicx}
\usepackage{makecell}
\usepackage{stfloats}
\usepackage{balance}
\usepackage{xcolor}
\usepackage{hyperref}
\hypersetup{hidelinks,pdfauthor={Fei Li, Haoxin Li, Kehong Liu, Yanzi Zhou, Weiguo Wu, Shiqiang Nie, Jinyu Wang, Song Liu},
  pdftitle={AscSpMM: Efficient SpMM on Ascend NPUs via Pipelined Vector-Cube Collaboration}}
\newcommand{\Description}[1]{}
\graphicspath{{figures/}}
\newcommand{\matA}{\boldsymbol{A}}
\newcommand{\matB}{\boldsymbol{B}}
\newcommand{\matC}{\boldsymbol{C}}
\newcommand{\Aindptr}{\ensuremath{\boldsymbol{A}_{\mathrm{indptr}}}\xspace}
\newcommand{\Acols}{\ensuremath{\boldsymbol{A}_{\mathrm{cols}}}\xspace}
\newcommand{\Adata}{\ensuremath{\boldsymbol{A}_{\mathrm{data}}}\xspace}
\newcommand{\Bpack}{\ensuremath{\boldsymbol{B}_{\mathrm{pack}}}\xspace}
\newcommand{\AscSpMM}{\textsc{AscSpMM}\xspace}

\title{\AscSpMM: Efficient SpMM on Ascend NPUs via Pipelined Vector--Cube Collaboration}

\author{Fei~Li, Haoxin~Li, Kehong~Liu, Yanzi~Zhou, Weiguo~Wu,
  Shiqiang~Nie, Jinyu~Wang, and~Song~Liu%
\thanks{This work has been submitted to the IEEE for possible publication. Copyright may be transferred without notice, after which this version may no longer be accessible.}%
\thanks{The authors are with the School of Computer Science and Technology, Xi'an Jiaotong University
  (e-mail: lifei@stu.xjtu.edu.cn; lhx20030110@stu.xjtu.edu.cn; 3125351031@stu.xjtu.edu.cn; 18737405312@stu.xjtu.edu.cn; wgwu@xjtu.edu.cn; shiqiang.nie@xjtu.edu.cn; jinyu.wang@xjtu.edu.cn; liusong@xjtu.edu.cn).}}

\begin{document}
\pagestyle{plain}
\bstctlcite{IEEEexample:BSTcontrol}

\maketitle
\thispagestyle{plain}

\begin{abstract}
Neural processing units (NPUs) provide high compute density and energy efficiency through specialized matrix engines and software-managed on-chip memory. However, their libraries primarily target dense computation. Sparse matrix–dense matrix multiplication (SpMM), a key kernel in graph analytics, sparse neural networks, and scientific computing, requires index-driven operand selection, indirect addressing, scattered memory accesses, and load balancing. These requirements conflict with the fixed shapes and contiguous layouts preferred by NPU matrix engines. GPU Tensor Core SpMM methods cannot be directly transferred to NPUs with decoupled vector and matrix engines. We present \AscSpMM, which separates irregular operand preparation from regular matrix computation through a parameterized Vector-CSR representation. The vector engine decodes indices and gathers and reorganizes dense operands, while the Cube engine performs matrix multiply–accumulate operations. A double-slot ring buffer enables dependency-aware asynchronous pipelining, while operand reuse and runtime-adaptive tiling reduce cross-engine stalls. On an Ascend 910B3, AscSpMM achieves geometric-mean speedups of 2.36× over GEMM, 19.92× over an existing SpMM implementation, and 5.82× over cuSPARSE on an A100. It reduces matrix-input storage and peak HBM footprint by 48.24\% and 22.98\% on average, while remaining competitive with specialized GPU Tensor Core SpMM kernels. The code is available at https://github.com/LfLab-AI/Ascend-NPU-SpMM.

\end{abstract}


\section{Introduction}\label{sec:introduction}

Sparse matrix--dense matrix multiplication (SpMM) is a fundamental kernel in scientific computing, graph analytics, recommender systems, and artificial intelligence \cite{davis2011suitesparse,bell2009spmv}. It supports neighborhood feature aggregation in graph neural networks (GNNs) \cite{ai2025neutronascend,kipf2017gcn}, embedding propagation over user--item graphs \cite{he2020lightgcn}, and the execution of sparse weights produced by model pruning \cite{gale2020sputnik,han2016deepcompression}. In principle, SpMM reduces both arithmetic work and sparse-matrix storage by skipping zero entries. In practice, sparse indices dynamically determine which rows of the dense operand participate in computation, introducing noncontiguous accesses, indirect addressing, metadata processing, and row-level load imbalance. Fewer multiply--accumulate operations therefore do not automatically produce proportional speedups. SpMM optimization must reconcile two conflicting requirements: sparsity favors fine-grained skipping, whereas high-throughput hardware depends on regular layouts, continuous operand supply, and sufficiently coarse computation \cite{yang2018design,bell2009spmv,hong2019aspt}.

GPUs have long dominated AI and scientific computing through massive parallelism and mature programming ecosystems. NPUs offer a more specialized approach to neural-network computation. Their matrix engines, dedicated data paths, and software-managed on-chip memories reduce general-purpose control and data-movement overhead for regular tensor workloads. This specialization provides a basis for high compute density and energy efficiency \cite{liao2021ascend,jouppi2017tpu}. NPUs have expanded from specialized inference accelerators into important platforms for large-scale training and inference \cite{jouppi2023tpuv4,liao2021ascend}. Broader operator support is essential to realizing this computational potential. However, current NPU software stacks remain centered on regular operators such as GEMM, convolution, and attention, while high-performance implementations of SpMM remain limited. Efficient SpMM is therefore important for extending NPU acceleration to graph analytics, sparse neural networks, and scientific computing.

Heterogeneous NPUs, represented by Google TPUs and Huawei Ascend NPUs, have been widely deployed in data centers and AI computing platforms \cite{jouppi2023tpuv4,liao2021ascend,jouppi2017tpu}. Ascend adopts the DaVinci architecture, which integrates matrix (Cube), vector, and scalar engines and supports a broad range of scenarios, from cloud training and inference to intelligent devices and autonomous driving \cite{liao2021ascend}. It has also been used in practical workloads such as GNNs \cite{ai2025neutronascend}, large language model training, and seismic simulation \cite{wang2025ascendseismic}. However, most high-performance SpMM studies still target GPUs \cite{yang2018design,gale2020sputnik,huang2020gespmm}, with recent work increasingly exploiting Tensor Cores \cite{chen2021vectorsparse,li2022magicube,okanovic2024smat,shi2025flashsparse}. These methods typically rely on intra-warp cooperation to integrate index decoding, scattered loading, shared-memory reorganization, and matrix-instruction invocation within a unified SIMT execution context. On Ascend NPUs, by contrast, the AI vector engines (AIV) and Cube engines (AIC) execute relatively independent instruction streams and use separate local memories. If data preparation relies on per-index Scalar loops, address calculation and control overhead limit operand supply. Vectorizing this index-driven preparation, however, requires meeting data-alignment and operand-layout constraints. In addition, the engines exchange prepared operands through global memory, introducing additional reads and writes and explicit synchronization overhead. These characteristics make it very difficult to implement high-performance SpMM on NPUs, and the architectural differences also mean that GPU-oriented SpMM designs cannot be directly transferred to this architecture.

We observe that the decoupled execution architecture of NPUs also provides a natural division of labor across processing stages for sparse tasks. The AIV is well suited to sparse-index processing and data reorganization, whereas the AIC provides high-throughput regular multiply--accumulate operations. However, exploiting this complementarity requires addressing three coupled challenges: constructing a sparse representation compatible with regular matrix computation, partitioning irregular data preparation and matrix computation across the two engines, and matching their producer--consumer rates with limited on-chip buffer capacity.

To this end, we present \AscSpMM, which co-designs sparse representation, task mapping, and cross-engine pipelining for decoupled NPUs. \AscSpMM introduces Vector-CSR, which organizes consecutive rows into row blocks matching Cube's computation granularity and uses shared active-column indices to link sparse values with the required dense operands. AIV gathers and reorganizes dense operands, while AIC performs the primary multiply--accumulate workload. A producer--consumer pipeline coordinates operand delivery, buffer reuse, and synchronization to overlap data preparation, data movement, and computation. Runtime-adaptive tiling further selects execution configurations based on active-column counts, matrix dimensions, and on-chip resources, adapting this cooperation to diverse sparse workloads.

The main contributions of this paper are as follows:

\begin{itemize}
\item We propose Vector-CSR, a sparse storage format that links index-driven data selection to regular operand loading through a shared sequence of active-column indices. The format constructs vector blocks according to the actual nonzero distribution and the computation granularity of Cube, without requiring rows to share a predefined sparsity pattern, thereby providing a unified execution mapping for structured and unstructured inputs.
\item We design a Vector--Cube cooperative dataflow that partitions work by processing stage. It assigns index-driven data preparation within the same SpMM task to AIV and maps the primary matrix multiply--accumulate workload to AIC. AIV exploits contiguous row segments and data sharing within row blocks to select rows and perform block-level reorganization, constructing regular operands for Cube so that the processing capabilities of both independent engines contribute to sparse computation.
\item We coordinate the AIV and AIC through a double-slot ring buffer, dependency-aware synchronization, early buffer release, operand reuse, prefetching, and runtime-adaptive tiling to reduce stalls across diverse sparse workloads.
\item We evaluate \AscSpMM on an Ascend 910B3 NPU. It achieves geometric-mean speedups of $2.36\times$ over GEMM and $19.92\times$ over NeutronSparse, while reducing matrix-input storage and peak HBM footprint by 48.24\% and 22.98\% on average. It also outperforms cuSPARSE by $5.82\times$ on average on a benchmark competitor A100 GPU, and delivers highly competitive performance with Sputnik and Magicube.
\end{itemize}

\section{Related Work}\label{sec:related-work}

\subsection{SpMM Optimization on NVIDIA GPUs}\label{sec:gpu-work}

GPU SpMM optimization has followed two main directions. CUDA Core implementations retain general CSR-like inputs and improve sparse-row partitioning, coalesced memory access, data reuse, and load balance. Merge-based SpMM, ASpT, Sputnik, GE-SpMM, and RoDe are representative examples \cite{yang2018design,hong2019aspt,gale2020sputnik,huang2020gespmm,pang2024rode}. These methods support general unstructured matrices, but their main multiply--accumulate workload remains on CUDA Cores and therefore cannot fully exploit the throughput of modern Tensor Cores.

A second line of work maps sparse computation to Tensor Cores through vectorized sparse formats, matrix reordering, or locally regular computation blocks. vectorSparse and Magicube exploit one-dimensional (1D) vector sparsity and index sharing \cite{chen2021vectorsparse,li2022magicube}; TC-GNN, SMaT, and DTC-SpMM construct or schedule regular tiles for general sparse matrices \cite{wang2023tcgnn,okanovic2024smat,fan2024dtcspmm}; and FlashSparse and Acc-SpMM further reduce redundant computation and improve pipeline efficiency \cite{shi2025flashsparse,zhao2025accspmm}. SSpMM combines sparse vector compression, transpose mapping, and register shuffling to support structured and unstructured SpMM across multiple generations of Tensor Cores \cite{xue2025sspmm}. Despite their different representations and schedules, these designs rely on GPU-specific warp cooperation, shared memory, and a unified SIMT execution context.

\subsection{SpMM Optimization on NPUs}\label{sec:npu-work}

Compared with the extensive body of GPU research, SpMM optimization on NPUs and dedicated AI accelerators remains relatively limited. Early systems such as EIE, Cambricon-X/S, SCNN, Eyeriss v2, ExTensor, SIGMA, and NVR improve sparse execution through compressed representations, sparse dataflows, flexible interconnects, dynamic task distribution, or address-generation support \cite{han2016eie,zhang2016cambriconx,zhou2018cambricons, parashar2017scnn,chen2019eyerissv2,hegde2019extensor, qin2020sigma,wang2025nvr}. These approaches largely rely on customized hardware paths and often target specific sparse DNN or tensor workloads. On Ascend NPUs, NeutronAscend optimizes sparse aggregation within an end-to-end GNN training pipeline \cite{ai2025neutronascend}. NeutronSparse directly targets SpMM by assigning matrix regions to AIV or AIC according to local sparsity \cite{ai2026neutronsparse}. However, regions mapped to AIC are still executed as regular 2D blocks, making efficiency sensitive to local nonzero density and partition balance.

Publicly available studies on NPU sparse computation therefore remain largely focused on system-level acceleration for GNNs, sparse neural networks, and specific models, with comparatively little attention devoted to fine-grained optimization of SpMM as an independently reusable primitive. The primary reason is a fundamental architectural mismatch: sparse indices require fine-grained metadata decoding, indirect addressing, and scattered data access, whereas NPU matrix engines are optimized for fixed-shape operands, contiguous layouts, and regular data supply. The scalar control path cannot readily sustain the operand supply required by high-throughput matrix computation through per-index processing, making operand preparation itself a key optimization target. \AscSpMM effectively addresses this challenge by constructing a fine-grained cross-engine cooperative data path from a CSR-like sparse representation to high-throughput matrix computation.

\section{Background}\label{sec:background}

\subsection{SpMM Computation and Sources of Irregularity}\label{sec:spmm-background}

Given a sparse matrix $\matA\in\mathbb{R}^{M\times K}$ and a dense matrix $\matB\in\mathbb{R}^{K\times N}$, SpMM produces $\matC\in\mathbb{R}^{M\times N}$ as
\begin{equation}
\matC_{i,j}=\sum_{k:\matA_{i,k}\neq0} \matA_{i,k}\matB_{k,j}.
\label{eq:spmm-element}
\end{equation}
SpMM reduces the $O(MKN)$ multiply--accumulate operations of GEMM to $O(\operatorname{nnz}(\matA)N)$. The reduction introduces irregularity because the summation set depends on the nonzero positions in each row of $\matA$. For every nonzero $(i,k)$, the implementation retrieves the column index $k$ and accesses the corresponding row of $\matB$; noncontiguous indices therefore create scattered memory accesses across global memory and caches. Variations in the number of nonzeros per row also translate directly into workload imbalance \cite{yang2018design,bell2009spmv}.

The irregularity is concentrated along $K$, while exploitable regularity remains along $N$: each selected element $\matA_{i,k}$ multiplies a contiguous segment $\matB_{k,0:N}$, and adjacent output rows may reuse the same rows of $\matB$. Scalar CSR preserves fine-grained sparsity but exposes little regular structure to matrix engines. 2D block formats are easier to map onto such engines, but underfilled blocks introduce zero-valued transfers and redundant multiply--accumulate operations. \AscSpMM seeks a middle ground by preserving index-based selection while reorganizing only the selected operands into regular matrices.

\subsection{NPU Execution Model and On-Chip Memory Hierarchy}\label{sec:npu-arch}

\begin{figure*}[!t]
\centering
\includegraphics[pagebox=cropbox,width=0.90\textwidth]{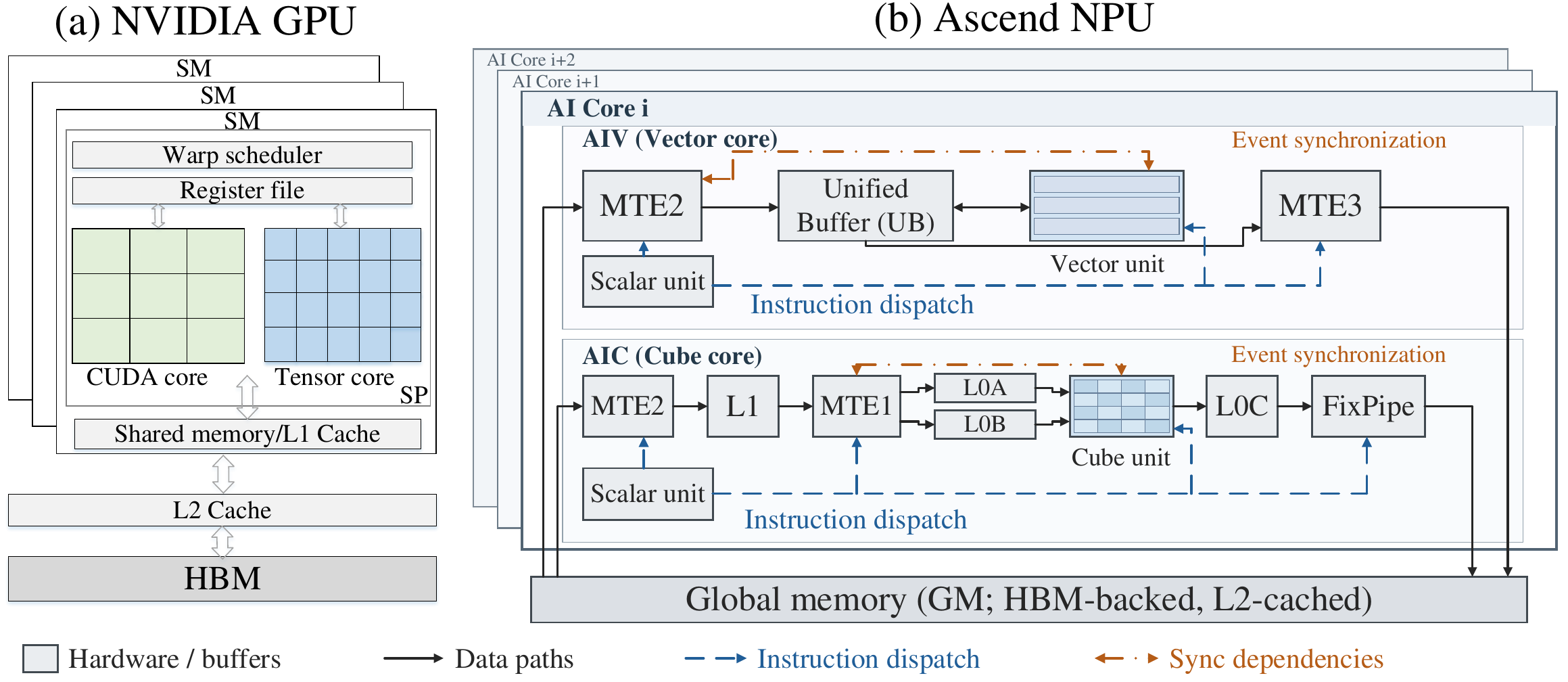}
\Description{Comparison of a GPU streaming multiprocessor with an Ascend NPU AI Core and their memory hierarchies.}
\caption{GPU and Ascend NPU compute architectures.}
\label{fig:architecture}
\end{figure*}

Modern GPUs and NPUs both include dedicated matrix units, but organize execution differently. As shown in Fig.~\ref{fig:architecture}(a), a GPU uses the streaming multiprocessor as its basic execution unit. CUDA Cores and Tensor Cores operate within a unified SIMT environment, allowing threads in the same block to coordinate address calculation, data loading, shared-memory reorganization, matrix-instruction invocation, and operand reuse. NPUs with decoupled execution instead assign control, vector processing, and matrix computation to relatively independent scalar, vector, and matrix engines. Each engine maintains its own instruction stream and local memory, so data movement, layout transformation, and execution order must be orchestrated explicitly by the compiler or operator implementation.

The Ascend NPU is a prime example of this decoupled architecture \cite{liao2021ascend,huaweicann850docs}. As shown in Fig.~\ref{fig:architecture}(b), AIV and AIC execute independent programs and use isolated local memories; intermediate data is exchanged through a global-memory (GM) workspace. AIV combines Scalar and Vector units, memory transfer engines, and a Unified Buffer (UB), making it suitable for index decoding, address generation, data selection, and reorganization. AIC performs regular matrix multiply--accumulate operations and result write-back through L1, L0A/L0B, Cube, L0C, and FixPipe. Data on the AIV path moves through GM, UB, Vector, and back to GM, whereas AIC operands move through GM, L1, and L0A/L0B before Cube execution, with results returned through L0C and FixPipe. memory transfer engine (MTE) 1, MTE2, Cube, and FixPipe can operate asynchronously, but synchronization events must follow the actual read--write dependencies.

This architecture imposes three constraints on SpMM. First, conventional CSR does not directly form the fixed-shape operands required by the Cube unit, whereas padding sparse data into 2D blocks wastes transfer and computation. Second, sparse column indices induce scattered accesses to rows of $\matB$; handling them individually on Scalar or issuing them directly from AIC can make address generation and layout transformation bottlenecks. Third, the preparation rate of AIV and the consumption rate of AIC vary with sparse row width, matrix dimensions, and tiling. Because cross-engine exchange passes through GM, mismatched rates create additional data-movement overhead, mutual stalls, and pipeline bubbles. An efficient design must therefore regularize the selected operands, partition work across AIV and AIC, and coordinate their execution.

\section{Methodology}\label{sec:methodology}

\begin{figure*}[!t]
\centering
\includegraphics[pagebox=cropbox,width=\textwidth]{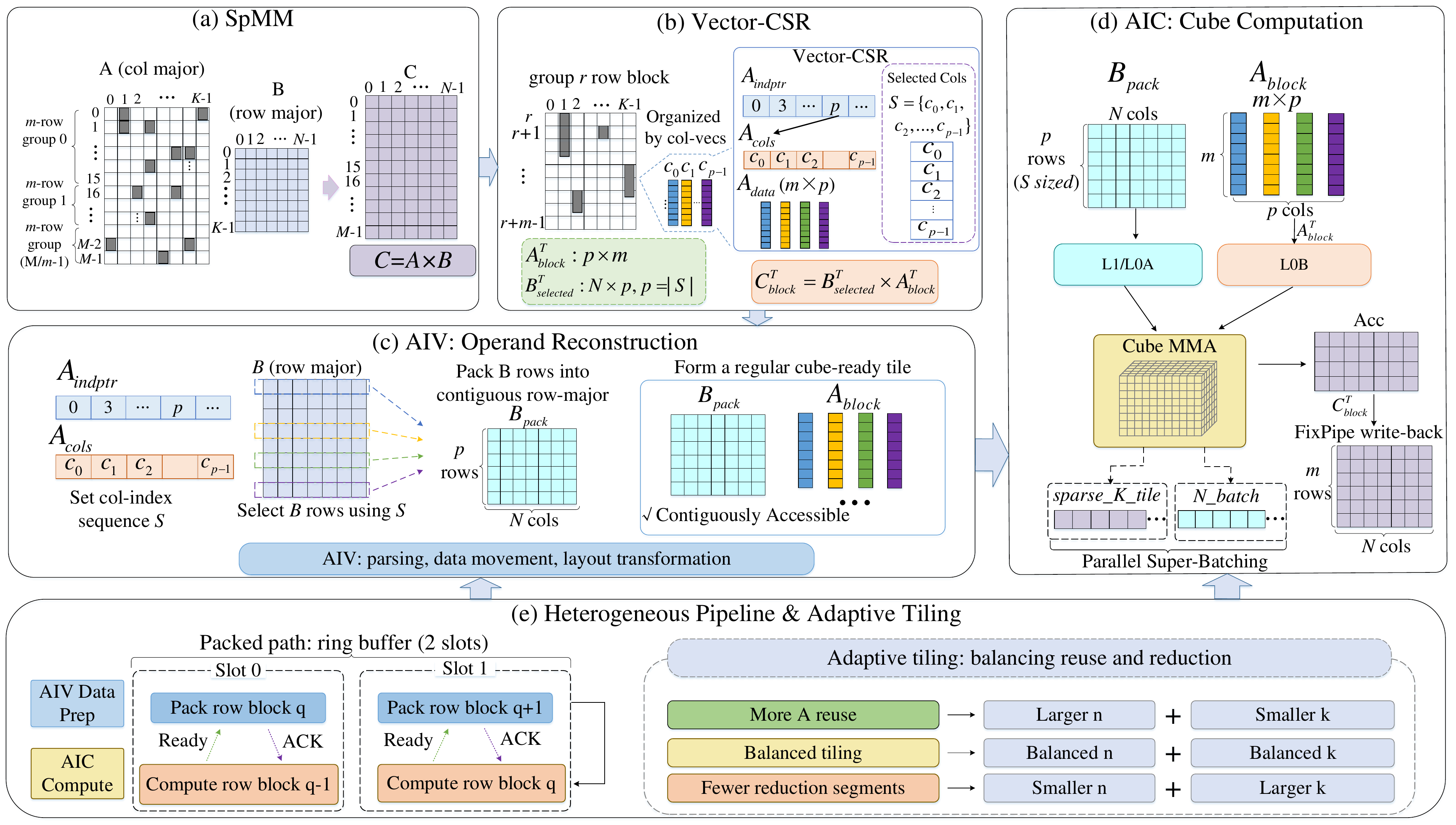}
\Description{Overview of the AscSpMM representation, operand packing, AIV--AIC cooperative computation, pipeline, and runtime tiling.}
\caption{Overview of \AscSpMM.}
\label{fig:overview}
\end{figure*}

\AscSpMM separates SpMM into sparse-operand organization, index-driven dense-operand packing, and regular Cube computation. As shown in Fig. \ref{fig:overview}, Vector-CSR groups $m$ consecutive rows of $\matA$ and records their active columns. AIV gathers the corresponding rows of $\matB$ into contiguous $\Bpack$ panels, which AIC consumes together with the Vector-CSR values as regular Cube operands. A ring-buffer pipeline overlaps operand preparation and matrix computation, while runtime tiling adapts their processing granularities to sparse-row width and on-chip resource limits.

\subsection{Vector-Block Sparse Representation: Vector-CSR}\label{sec:vector-csr}

Conventional CSR preserves fine-grained sparsity but associates each nonzero with an independent index lookup and dense-operand access, exposing insufficient regularity to fixed-shape Cube operations. 2D block formats improve execution regularity but may introduce substantial padding for irregular sparsity. 

\AscSpMM therefore introduces Vector-CSR, a vector-block sparse storage format designed for the NPU backend. The format adopts a 1D column vector as its organizing granularity, sharing active-column indices across multiple output rows while retaining column-level sparse selection at a finer granularity than 2D matrix blocks. Its basic storage unit is a vertical $m\times1$ vector block along the $M$ dimension of $\matA$. Here, $m$ is a backend-dependent vector granularity selected according to the target 1D vector-block size and the operand shapes of the underlying hardware matrix instructions. In our Ascend NPU implementation, $m$ is set to 16 to match the output-row granularity of the Cube primitive; other backends can use different values to match their hardware matrix primitives. \AscSpMM groups every $m$ consecutive rows along the $M$ dimension into a row block. Within a row block, a column is considered active if it contains at least one nonzero entry across these $m$ rows, and a corresponding vector block is created for that column. The $m$ row positions in a vector block share a single column index. Multiple active columns are arranged along the $K$ dimension, forming a local matrix with a fixed output-row dimension of $m$. When the final row block contains fewer than $m$ rows, the invalid positions are zero-padded, and only valid output rows are retained during write-back.

\begin{figure}[!t]
\centering
\includegraphics[pagebox=cropbox,width=0.88\columnwidth]{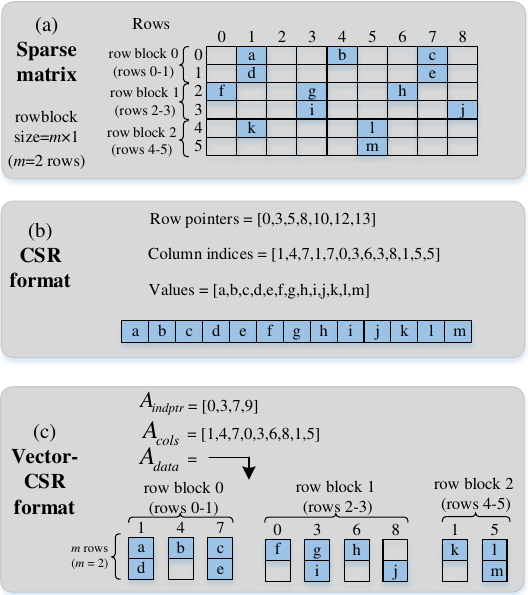}
\Description{Example converting sparse matrix rows from CSR into Vector-CSR blocks.}
\caption{CSR and Vector-CSR formats.}
\label{fig:vector-csr}
\end{figure}

As shown in Fig.~\ref{fig:vector-csr}, Vector-CSR consists of three arrays: $\Aindptr$, $\Acols$, and $\Adata$. The length of \Aindptr is one greater than the number of row blocks, and its entries record the cumulative boundaries of row blocks in the sequence of active vector blocks. The number of vector blocks in the $q$-th row block is $p_q=\Aindptr[q+1]-\Aindptr[q]$.

Let $P=\sum_q p_q$ denote the total number of active vector blocks
stored for the entire matrix. The length of \Acols is $P$, and the
entry \Acols[$t$] records the original column index associated with
the $t$-th vector block. Following the same order, \Adata stores $P$
vector blocks, each containing $m$ contiguously stored values.
Consequently, \Aindptr determines how many vectors belong to each row
block, \Acols identifies the column associated with each vector, and
\Adata stores the values of that active column across the $m$ row
positions.

Compared with scalar-granularity CSR, Vector-CSR organizes the values
of $\matA$ into fixed-height regular vector blocks and allows the $m$
row positions associated with the same active column in a row block to
share a single column index. The resulting degree of index compression
depends on the column-wise clustering of nonzeros within each row block.
Vector-CSR can represent both structured and unstructured sparsity: any
column containing at least one nonzero within a row block is recorded as
an active vector, while highly clustered nonzeros lead to higher vector
occupancy and greater index sharing. We therefore do not characterize
Vector-CSR as an unconditionally more compact storage format for every
sparse distribution. Instead, it serves as a backend execution
representation that bridges the logical sparse structure and the regular
operands required by the Cube unit. Vector-CSR is particularly effective for matrices generated by vector-wise pruning, block-structured pruning, structured attention
patterns, and other sparsity constraints. In such workloads, nonzeros
often occur in small, fixed-shape structures, resulting in relatively
high active-vector occupancy and predictable local distributions that
facilitate regular operand construction with modest metadata overhead.

\subsection{Vector--Cube Cooperative Computation}\label{sec:cooperative}

Vector-CSR stores $\matA$ as contiguous fixed-length vectors, but SpMM still requires index-driven gathering of the corresponding rows of $\matB$. On Ascend NPUs, the absence of GPU-style warp-centric cooperation makes scattered address generation, loading, and layout reorganization inefficient within a single execution context; implementing these operations as per-index Scalar loops would further serialize data preparation. \AscSpMM therefore separates regular operand loading from irregular data preparation: \Adata is delivered directly to AIC as a regular matrix operand, while AIV resolves the active-column indices and gathers and packs the selected rows of $\matB$ into contiguous tiles on the fly.

For the $q$-th row block, let $\matA^{(q)}\in\mathbb{R}^{m\times p_q}$ denote the local matrix formed by its $p_q$ active vector blocks, and let $\matB^{(q)}\in\mathbb{R}^{p_q\times N}$ denote the rows selected from $\matB$ according to the same active columns. The local computation is $\matC^{(q)}=\matA^{(q)}\matB^{(q)}$.

AIV first reads \Aindptr to determine the vector range of the current row block and then obtains the active-column indices from \Acols. For each production batch along the $N$ dimension, referred to as a packed panel, AIV partitions the active-vector range into subtiles and processes them sequentially. For each active vector, it transfers a contiguous segment of the corresponding row of $\matB$ from GM to UB. It then writes these row segments into \Bpack in active-vector order. Consequently, the $t$-th row of \Bpack is aligned exactly with the $t$-th vector block in \Adata. Although the selected source rows may be scattered along $K$, each row segment of $\matB$ remains contiguous along the $N$ dimension. This property allows AIV to replace element-wise transfers with longer block transfers.

The $\matA$ operand does not require the preceding index-driven reorganization. Each $m$-element vector block in \Adata is stored contiguously, allowing AIC to load the current tile along the active-column dimension and directly construct the regular operand $\matA^{(q)}$, or equivalently its transpose $\left(\matA^{(q)}\right)^{\mathsf{T}}$. To match the left- and right-operand paths of Cube, \AscSpMM evaluates the multiplication in the following transposed but mathematically equivalent form:
\begin{equation}
{\matC^{(q)}}^{\mathsf{T}}={\matB^{(q)}}^{\mathsf{T}}{\matA^{(q)}}^{\mathsf{T}}.
\label{eq:transposed-local-spmm}
\end{equation}

According to Eq.~\ref{eq:transposed-local-spmm}, AIC routes \Bpack through GM, L1, and L0A as Cube's left operand, while routing \Adata through GM, L1, and L0B as the right operand. Partial sums are accumulated in L0C. \Bpack is stored in row-major order. The AIC represents $\left(\matB^{(q)}\right)^{\mathsf{T}}$ through the source-stride configuration and the ND-to-fractal data-movement path, avoiding an explicit full transpose in GM. When $p_q$ exceeds the current sparse-$K$ tile, AIC loads $\matA$ and $\matB$ in multiple tiles and completes the reduction through successive Cube invocations.

During execution, AIV gathers the rows of $\matB$ selected by the active-column indices, reorganizes them in UB, and writes a contiguous $\Bpack$ tile to the GM workspace. After AIV signals that the tile is ready, AIC loads $\Bpack$ and $\Adata$ through L1 and L0A/L0B, executes the Cube operations, and writes the valid output tile through FixPipe. Synchronization is inserted only at producer--consumer and buffer-reuse dependencies, allowing otherwise independent data-transfer and computation stages to proceed asynchronously.

\subsection{Ring-Buffer Pipelining and Runtime-Adaptive Tiling}\label{sec:pipeline}

\begin{figure}[!t]
\centering
\includegraphics[pagebox=cropbox,width=\columnwidth]{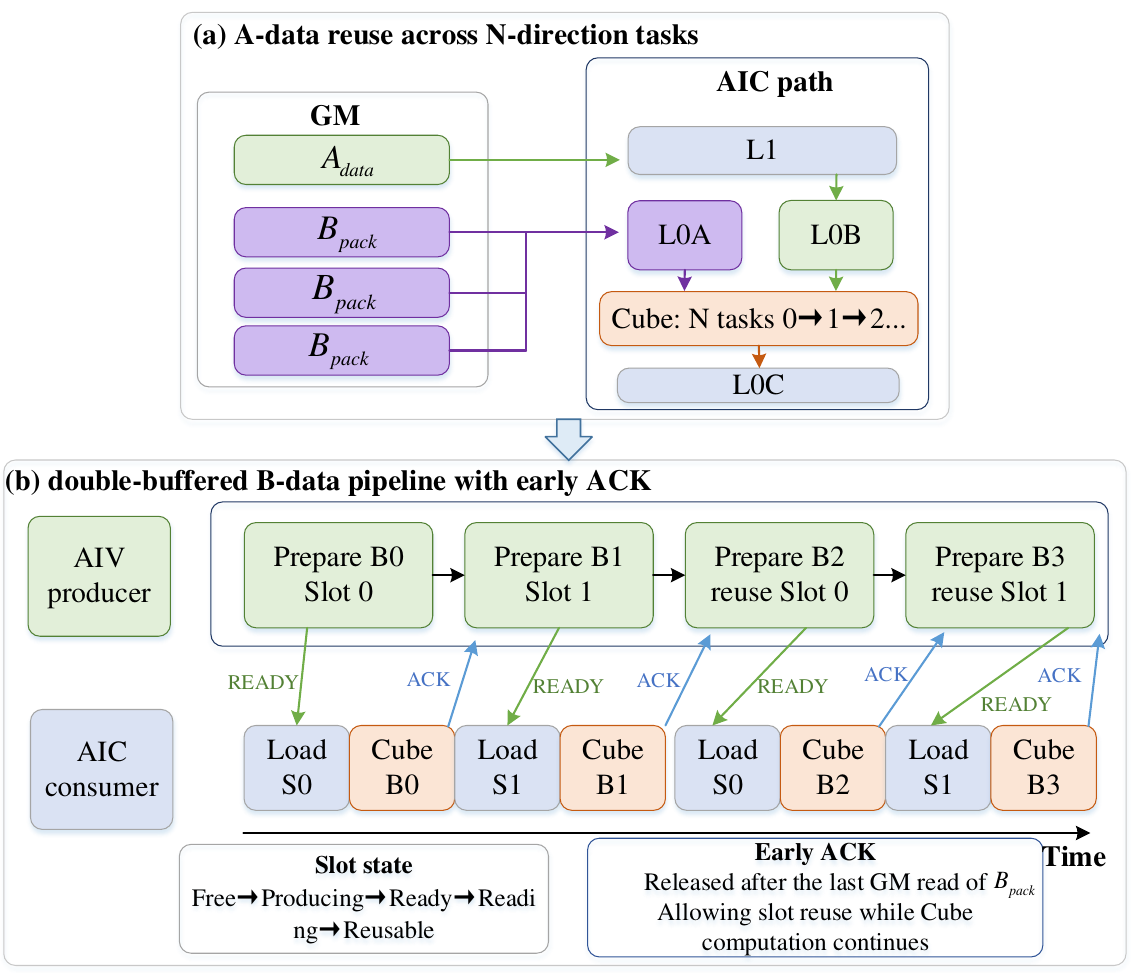}
\Description{$\matA$-data reuse and a double-buffered producer--consumer pipeline between AIV and AIC.}
\caption{Reuse of $\matA$ and the double-slot AIV--AIC pipeline.}
\label{fig:pipeline}
\end{figure}

\textbf{Ring-Buffer Pipelining.} In the dataflow of Section~\ref{sec:cooperative}, AIV and AIC become effectively serialized if AIC must wait until AIV completes all data preparation for the current task. The preparation time on AIV depends on the number of active vectors, the access pattern of the selected rows of $\matB$, and the $N$ dimension, whereas the execution time on AIC is governed primarily by tile sizes and Cube utilization. Their processing rates therefore cannot remain consistently matched across workloads. To reduce mutual stalls, \AscSpMM combines $\matA$ operand reuse with a double-slot ring buffer, overlapping the preparation of $\matB$ with Cube computation, as illustrated in Fig.~\ref{fig:pipeline}.

For a row block that fits within one active-vector tile, the $\matA$ operand is independent of the current interval along the $N$ dimension. \AscSpMM loads it into L0B once and updates only the $\matB$ operand as the output-column range advances. This design reduces repeated transfers of $\matA$ from GM to L1 and L0B and amortizes its layout-transformation cost across multiple Cube tasks. Because $\matB$ must be updated for each $N$-dimensional window, it becomes the primary object transferred through the pipeline between AIV and AIC. As shown in Fig.~\ref{fig:pipeline}(b), \AscSpMM uses two cyclically reused GM buffer slots to exchange the packed panels of each row block. After reorganizing a batch of $\matB$ into one slot, AIV issues a READY signal. AIC sends an ACK immediately after completing the final GM read from that slot. The subsequent L1-to-L0A transfer, Cube computation, and FixPipe write-back do not access that slot, allowing AIV to overwrite the released slot and prepare the next batch while AIC continues computing. This early-acknowledgment mechanism releases each slot at the end of its actual data dependency and thereby enlarges the overlap window between AIV preparation and AIC computation.

\textbf{Runtime-Adaptive Tiling.} AIV and AIC have different processing-granularity requirements. In the packed path, AIV prepares wide panels along the $N$ dimension to increase contiguous transfer lengths and amortize index processing and synchronization. AIC partitions these panels into computation tiles that fit in on-chip memory. Wider output tiles improve $\matA$ reuse but increase the storage required for $\matB$ and the output. Larger tiles along the active-vector dimension reduce the number of reduction segments but also require more on-chip space. \AscSpMM therefore configures the two engines' processing granularities separately and selects them adaptively at runtime according to matrix dimensions and sparsity structure.

Let $n$ and $k$ denote the numbers of output columns and active vectors in an AIC computation tile, respectively, and let $P_{\mathrm{pack}}$ denote the AIV preparation width. In the base path, an FP16 $\matB$ tile occupies $2nk$ bytes, so the 64\,KiB capacity of L0A requires $nk\le32768$. Subject also to the capacity constraints of $\matA$, the output tile, and the UB panel, \AscSpMM precompiles the following base tile configurations:
\[
\begin{aligned}
(n,k)\in\mathcal{C}_{\mathrm{base}}=\{&(1024,32),(512,64),(256,128),\\
&(128,256),(16,2048)\}.
\end{aligned}
\]
The host scans the row pointers to obtain the maximum row-block length, the average length of nonempty row blocks, and the task count. It combines these statistics with $M,N,K$ to select a base configuration. In some short-row configurations, AIC reads the required rows of $\matB$ directly, avoiding packing and cross-engine handoff. Packed configurations use compatible preparation widths $P_{\mathrm{pack}}\in\{512,1024,2048\}$.

For very short row blocks, tile-size adjustment alone offers limited amortization of the overhead of executing each row block separately. \AscSpMM therefore fuses 2, 4, or 8 consecutive row blocks into one task when all are nonempty and the capacity constraints are satisfied. AIV constructs a block-diagonal $\matA$ panel and the corresponding $\matB$, allowing AIC to process multiple row blocks together and amortize data preparation, computation setup, and write-back. The host selects the fusion group size according to on-chip capacity, the fraction of fusible row blocks, and the resulting per-core workload. Incomplete or ineligible groups are processed row block by row block.

For wide outputs, traversing all output columns for each row block expands the $\matB/\matC$ access range across consecutive tasks, limiting local reuse. \AscSpMM supports traversal in slabs of 1,024 columns, processing the row blocks within the current slab before advancing to the next. This can narrow the $\matB/\matC$ access range and improve local reuse at the cost of reloading $\matA$ across slabs. The host selects this path when $N>1024$, the original $\matB$ access range and per-core footprint are large, the estimated slab working set fits within the cache budget, and sufficient tasks are available. Eligible short-row fusion takes precedence; otherwise, the host selects slab traversal or a base configuration.

\section{Evaluation}\label{sec:evaluation}

\subsection{Experimental Setup}\label{sec:setup}

We evaluate \AscSpMM on an Ascend 910B3 NPU server and an NVIDIA A100 GPU server. Table~\ref{tab:platforms} summarizes their hardware and software configurations. The two platforms represent a matrix--vector decoupled NPU and a SIMT/Tensor Core GPU, respectively, and provide comparable peak FP16 matrix throughput. We use the A100 results as a cross-platform reference to assess \AscSpMM against mature GPU SpMM implementations.

\begin{table}[!t]
\caption{Experimental platforms and software stacks.}
\label{tab:platforms}
\centering
\footnotesize
\begin{tabularx}{\columnwidth}{@{}lXX@{}}
\toprule
\textbf{Configuration} & \textbf{Ascend 910B3} & \textbf{NVIDIA A100} \\
\midrule
Accelerator & Ascend 910B3 & A100-SXM4-40GB \\
Peak FP16 FLOPS & 313 TFLOPS \cite{huaweiatlas800ta2whitepaper} & 312 TFLOPS \cite{nvidia2020a100} \\
HBM bandwidth & 1,600 GB/s \cite{huaweiatlas800ta2whitepaper} & 1,555 GB/s \cite{nvidia2020a100} \\
Host processor & Kunpeng 920 5250; 192 cores & Intel Xeon Platinum 8173M; 56 cores \\
OS & openEuler 22.03 LTS-SP4; Linux 5.10.0 & Ubuntu 22.04.5 LTS; Linux 6.8.0 \\
Software stack & CANN 8.5.0; CCE clang 15.0.5, torch\_npu 2.9.0; MindSpore 2.7.2~\cite{mindspore2026docs}; GraphLearning 0.2 & CUDA 12.4;  cuSPARSE 12.3; cuBLAS 12.4 \\
\bottomrule
\end{tabularx}
\end{table}

\noindent\textbf{Workloads.}
The controlled evaluation covers $M=N=K\in\{64,256,512,1024,2048,4096,8192,16384\}$ and six sparsity levels---90\%, 95\%, 98\%, 99\%, 99.5\%, and 99.9\%---for a total of 48 configurations. The controlled evaluation uses matrices with 1D structured sparsity, generated by replacing each nonzero scalar in a randomly generated sparse matrix with a dense $16\times1$ vector. These workloads isolate the effects of matrix size and sparsity. We additionally evaluate 14 unstructured sparse matrices, comprising 13 matrices from the SuiteSparse Matrix Collection~\cite{davis2011suitesparse} and the ogbn-proteins adjacency matrix: pdb1HYS, cant, consph, pwtk, rail4284, web-Stanford, web-Google, pkustk12, x104, boneS01, PR02R, degme, Delor338K, and ogbn-proteins. They span scientific-computing problems, Web graphs, and graph workloads with diverse matrix shapes and nonzero distributions. For each matrix, we independently vary the dense feature width over $N\in\{128,256,512,1024,2048\}$ while retaining its row and column dimensions and sparsity pattern, yielding 70 configurations. We also evaluate end-to-end inference with an unpruned two-layer GCN of hidden width 256 on CiteSeer, Cora, Pubmed, Wiki-CS, and ogbn-arxiv in TorchNPU and MindSpore.

\noindent\textbf{Baselines.}
The same-platform baselines are Ascend GEMM (AscGEMM) \cite{huawei2026cblas}, NeutronSparse \cite{ai2026neutronsparse}, vecSpMM, and vec-cubeSpMM. Ascend GEMM is the BLAS implementation in the Ascend software stack and identifies the break-even point between sparse and dense execution. NeutronSparse assigns matrix regions to Cube or Vector according to local sparsity instead of using a CSR-like compressed representation. vecSpMM uses the same Vector-CSR representation as \AscSpMM but executes the entire computation on the Vector unit, measuring the benefit of mapping regular multiply--accumulate operations to Cube. vec-cubeSpMM retains Vector--Cube cooperation but disables the double-slot buffer and asynchronous cross-engine pipeline, isolating the benefit of pipeline overlap.

The A100 baselines include cuSPARSE \cite{nvidia2024cusparse}, the CUDA Core-based Sputnik \cite{gale2020sputnik}, and the Tensor Core-based Magicube \cite{li2022magicube}; cuBLAS serves as their dense-computation reference. All methods use identical $M$, $N$, and $K$ dimensions and sparsity levels. Unsupported configurations are reported as N/A.

\noindent\textbf{Measurement.}
We use Ascend \textit{msprof} \cite{huaweicann850docs} to measure device-side kernel execution time and peak HBM footprint. Each configuration is executed 21 times; after discarding the first run as warm-up, we report the mean of the remaining 20 runs. Effective throughput is $2\cdot\operatorname{nnz}(\matA)N/t$, where $t$ is the kernel execution time, and speedup is the baseline runtime divided by the runtime of \AscSpMM. For the controlled-matrix throughput plot, we estimate $\operatorname{nnz}(\matA)\approx MK(1-s/100)$ from the configured sparsity $s$ (in percent).

\subsection{Overall Performance}\label{sec:performance}

\begin{figure*}[!t]
\centering
\includegraphics[pagebox=cropbox,width=0.75\textwidth]{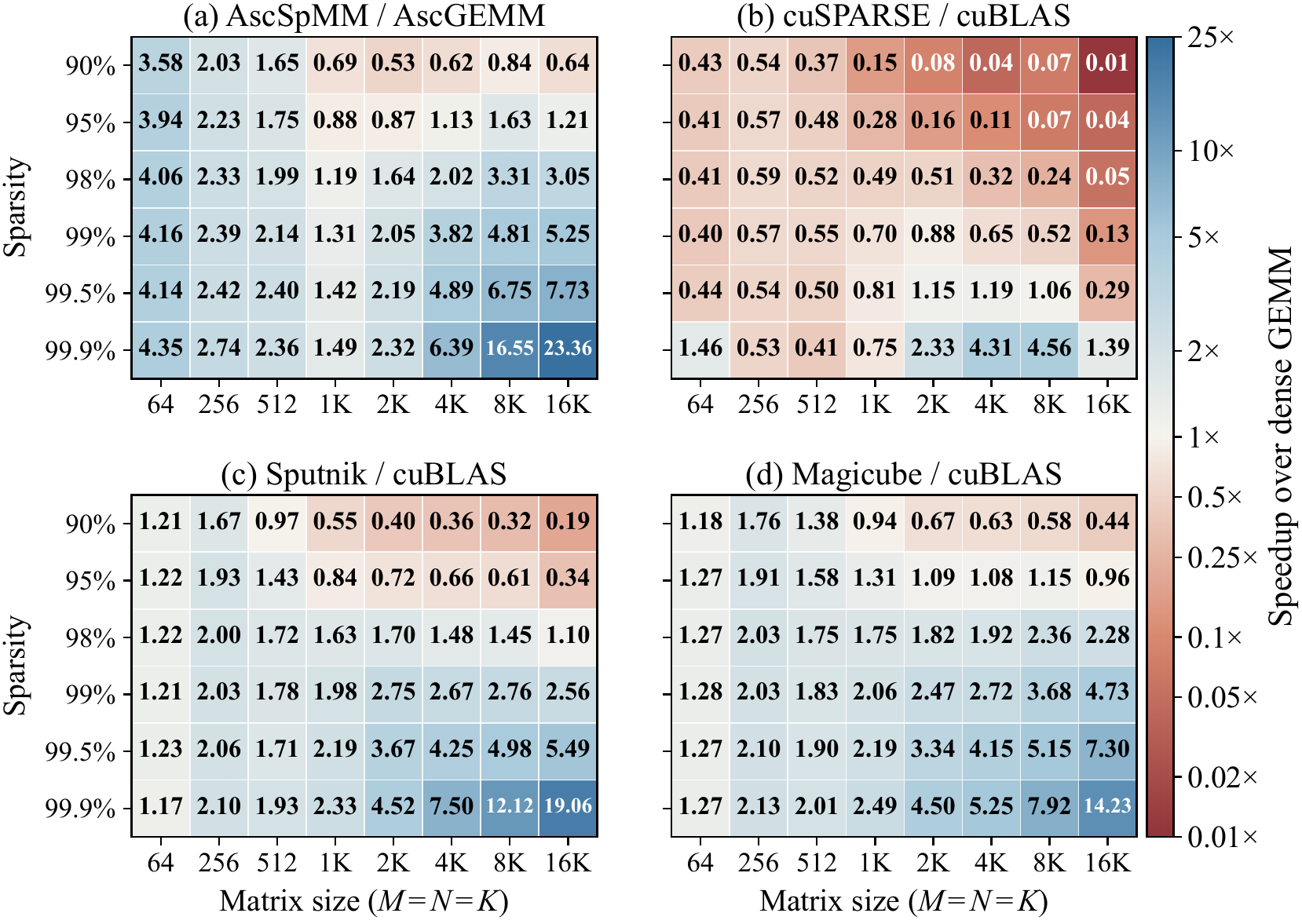}
\Description{Four heat maps of sparse-operator speedups over dense GEMM across matrix sizes and sparsity levels.}
\caption{Sparse-operator speedups over same-platform dense GEMM across 48 controlled workloads.}
\label{fig:dense-speedups}
\end{figure*}

\textbf{Performance relative to dense GEMM baselines.}\newline Fig.~\ref{fig:dense-speedups} (a)--(d) compare each sparse operator with the dense GEMM implementation on its respective NPU or GPU platform. \AscSpMM outperforms AscGEMM in 41 of the 48 configurations, with a geometric-mean speedup of $2.36\times$. For small matrices, dense GEMM underutilizes Cube, so the computation eliminated by sparsity can offset sparse-processing overhead even at 90\%--95\% sparsity. At $1\mathrm{K}^3$ and larger, dense GEMM achieves higher utilization, while \AscSpMM additionally incurs active-column decoding, \Bpack construction, and GM-based AIV--AIC data exchange. Some configurations therefore remain below the $1\times$ break-even point at 90\% sparsity. At sparsity levels of 98\% and above, however, \AscSpMM consistently outperforms AscGEMM across all tested sizes. Thus, its benefit depends on whether the reduction in Cube multiply--accumulate operations offsets regular-operand construction and cross-engine delivery costs. In addition to execution-time improvements, sparse computation also reduces matrix storage and peak HBM consumption relative to dense GEMM, as further discussed in Section \ref{sec:memory}.

Using dense GEMM on each platform as the reference, the geometric-mean speedups of \AscSpMM, cuSPARSE, Sputnik, and Magicube are $2.36\times$, $0.39\times$, $1.62\times$, and $1.94\times$, respectively, across all 48 configurations. cuSPARSE outperforms cuBLAS in only 8 of the 48 configurations because index processing, indirect access, and low arithmetic intensity often offset the reduction in computation. Sputnik and Magicube achieve more stable gains through sparsity-aware partitioning and Tensor Core mapping, respectively. Despite exchanging data between Vector and Cube through GM, \AscSpMM achieves a higher geometric-mean speedup over dense execution than these GPU-specialized kernels. This comparison does not imply higher absolute NPU throughput; rather, it shows that Vector-CSR and producer--consumer pipelining reduce the overhead of decoupled execution to a level comparable to mature GPU-specific implementations.

\begin{figure}[!t]
\centering
\includegraphics[pagebox=cropbox,width=1.0\columnwidth]{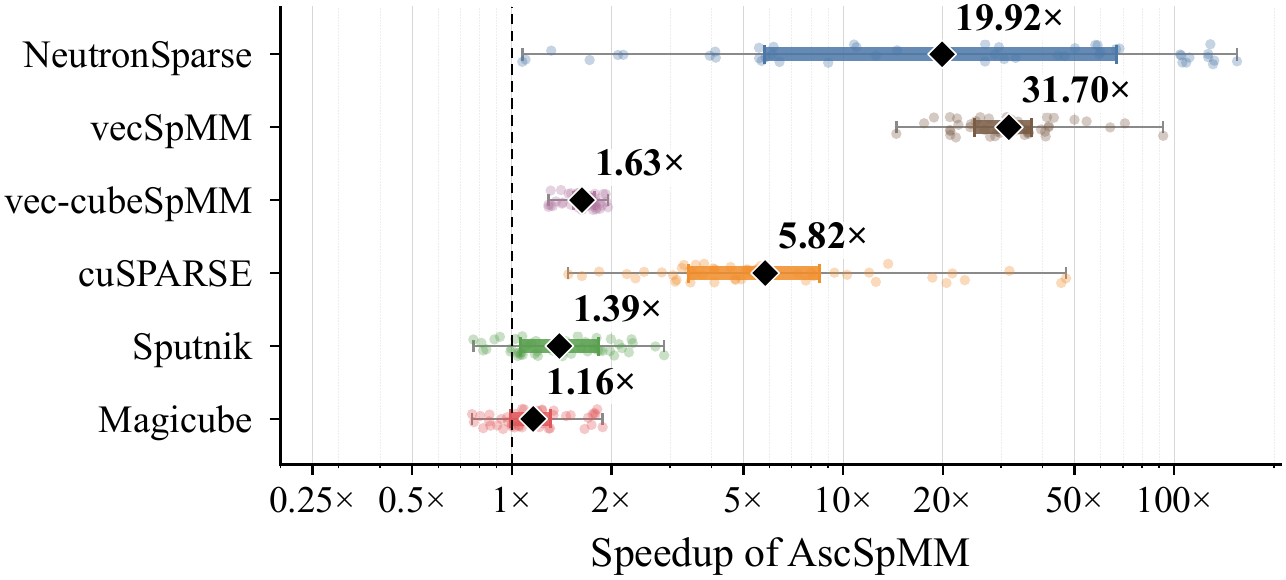}
\Description{Distributions of AscSpMM speedups over six sparse baselines.}
\caption{\AscSpMM speedups over six sparse baselines, each evaluated on 48 workloads; thin and thick lines show the minimum--maximum and interquartile ranges, respectively.}
\label{fig:sparse-speedups}
\end{figure}

\begin{table}[!t]
\caption{\AscSpMM speedups over sparse baselines across 48 workloads.}
\label{tab:sparse-baselines}
\centering
\footnotesize
\setlength{\tabcolsep}{3.2pt}
\begin{tabular}{@{}ccccc@{}}
\toprule
\textbf{Baseline} & \textbf{Cases} & \textbf{Wins} & \textbf{GM Speedup} & \textbf{Range ($\times$)} \\
\midrule
vecSpMM & 48 & 48 & $31.70\times$ & 14.49--92.67 \\
vec-cubeSpMM & 48 & 48 & $1.63\times$ & 1.29--1.95 \\
NeutronSparse & 48 & 48 & $19.92\times$ & 1.08--154.78 \\
cuSPARSE & 48 & 48 & $5.82\times$ & 1.48--47.10 \\
Sputnik & 48 & 40 & $1.39\times$ & 0.77--2.88 \\
Magicube & 48 & 35 & $1.16\times$ & 0.76--1.88 \\
\bottomrule
\end{tabular}
\end{table}

\textbf{Performance relative to sparse baselines and ablated implementations.} Fig.~\ref{fig:sparse-speedups} and Table~\ref{tab:sparse-baselines} compare \AscSpMM with NeutronSparse, vecSpMM, vec-cubeSpMM, and three GPU sparse kernels under identical matrix configurations.

\AscSpMM outperforms vecSpMM in all 48 configurations by a geometric mean of $31.70\times$. This result shows that Vector is effective for index decoding and data reorganization but lacks the throughput required for the primary multiply--accumulate workload when $N$ is large; mapping regularized operands to Cube is therefore essential. \AscSpMM also outperforms vec-cubeSpMM in all configurations by a geometric mean of $1.63\times$. Because both use the same representation and Vector--Cube task partitioning, this comparison measures the benefit of the double-slot ring buffer, early acknowledgment, and asynchronous cross-engine pipeline. Without these mechanisms, \Bpack construction and Cube computation are serialized; \AscSpMM overlaps them through double buffering and asynchronous execution. Relative to NeutronSparse, \AscSpMM achieves a geometric-mean speedup of $19.92\times$ and is faster in all 48 configurations. NeutronSparse assigns matrix regions independently to Vector or Cube, whereas \AscSpMM integrates the two engines into a producer--consumer dataflow that reduces redundancy through its sparse representation, task mapping, and execution pipeline.

\AscSpMM is faster than cuSPARSE on the A100 in all 48 configurations, with a geometric-mean speedup of $5.82\times$. Its geometric-mean speedups over Sputnik and Magicube are $1.39\times$ and $1.16\times$. The ranges are $0.77\times$--$2.88\times$ against Sputnik and $0.76\times$--$1.88\times$ against Magicube, showing that the relative advantage depends on the workload. These results demonstrate that \AscSpMM brings sparse computation on decoupled NPUs into the performance range of mature GPU implementations and achieves a higher geometric-mean performance than both specialized GPU kernels on this benchmark.

\textbf{Performance on unstructured matrices.} The controlled matrices isolate the effects of matrix size and sparsity, whereas the unstructured matrices evaluate adaptability to realistic nonzero distributions, row-length imbalance, and non-square structures. Table~\ref{tab:suitesparse} reports each matrix's structural characteristics, the execution time of \AscSpMM, and its speedups over the applicable baselines at $N=512$.

\begin{table*}[!t]
\caption{Execution time and speedups on 14 unstructured matrices with $N=512$.}
\label{tab:suitesparse}
\centering
\footnotesize
\setlength{\tabcolsep}{3.2pt}
\renewcommand{\arraystretch}{1.05}
\begin{tabular*}{\textwidth}{@{\extracolsep{\fill}}lcccccccc@{}}
\toprule
\multirow{2}{*}{\textbf{Matrix}} &
\multirow{2}{*}{\textbf{Dimensions}} &
\multirow{2}{*}{\makecell{\textbf{Sparsity}\\\textbf{(\%)}}} &
\multirow{2}{*}{\makecell{\textbf{Avg.}\\\textbf{nnz/row}}} &
\multirow{2}{*}{\makecell{\textbf{AscSpMM}\\\textbf{Time (ms)}}} &
\multicolumn{4}{c}{\textbf{Speedup}} \\
\cmidrule(lr){6-9}
& & & & & \textbf{NeutronSparse} & \textbf{cuSPARSE} & \textbf{Magicube} & \textbf{Sputnik} \\
\midrule
pdb1HYS & $36.4\mathrm{K}^{2}$ & 99.672 & 119.3 & 0.532 & $14.38\times$ & $6.88\times$ & $0.96\times$ & $1.44\times$ \\
cant & $62.5\mathrm{K}^{2}$ & 99.897 & 64.2 & 0.594 & $18.14\times$ & $5.50\times$ & $1.48\times$ & $1.32\times$ \\
consph & $83.3\mathrm{K}^{2}$ & 99.914 & 72.1 & 0.688 & $19.73\times$ & $7.30\times$ & $1.29\times$ & $1.58\times$ \\
pwtk & $217.9\mathrm{K}^{2}$ & 99.976 & 52.9 & 1.430 & $20.66\times$ & $6.84\times$ & $1.37\times$ & $1.63\times$ \\
rail4284 & $4.3\mathrm{K}{\cdot}1.1\mathrm{M}$ & 99.760 & 2634.0 & 1.426 & $72.47\times$ & $43.07\times$ & $1.31\times$ & $1.29\times$ \\
web-Stanford & $281.9\mathrm{K}^{2}$ & 99.997 & 8.2 & 1.634 & $12.81\times$ & $1.43\times$ & $1.94\times$ & $1.01\times$ \\
web-Google & $916.4\mathrm{K}^{2}$ & 99.999 & 5.6 & 4.268 & $15.77\times$ & $1.31\times$ & $1.26\times$ & $0.83\times$ \\
pkustk12 & $94.6\mathrm{K}{\cdot}94.7\mathrm{K}$ & 99.916 & 79.3 & 0.826 & $18.93\times$ & $7.44\times$ & $1.26\times$ & $1.59\times$ \\
x104 & $108.4\mathrm{K}^{2}$ & 99.926 & 80.4 & 0.967 & $18.18\times$ & $7.52\times$ & $1.28\times$ & $1.58\times$ \\
boneS01 & $127.2\mathrm{K}^{2}$ & 99.966 & 43.4 & 0.715 & $23.23\times$ & $6.19\times$ & $1.38\times$ & $1.48\times$ \\
PR02R & $161.1\mathrm{K}^{2}$ & 99.968 & 50.9 & 0.974 & $22.15\times$ & $7.05\times$ & $1.43\times$ & $1.68\times$ \\
degme & $185.5\mathrm{K}{\cdot}659.4\mathrm{K}$ & 99.993 & 43.5 & 2.071 & $35.65\times$ & $3.22\times$ & $1.69\times$ & $1.24\times$ \\
Delor338K & $343.2\mathrm{K}{\cdot}887.1\mathrm{K}$ & 99.999 & 12.4 & 1.899 & $18.50\times$ & $1.93\times$ & $0.84\times$ & $1.19\times$ \\
ogbn-proteins & $132.5\mathrm{K}^{2}$ & 99.775 & 298.5 & 3.672 & $18.11\times$ & $12.33\times$ & $1.13\times$ & $1.75\times$ \\
\bottomrule
\end{tabular*}
\end{table*}

At $N=512$, \AscSpMM outperforms NeutronSparse and cuSPARSE on all 14 matrices, with geometric-mean speedups of $21.02\times$ and $5.58\times$, respectively. Its geometric-mean speedups over Sputnik and Magicube are $1.37\times$ and $1.30\times$, with wins on 13 and 12 matrices, respectively. For pdb1HYS, cant, consph, pwtk, pkustk12, x104, and ogbn-proteins, the average row lengths range from 52.9 to 298.5 nonzeros. These longer local reductions provide more work over which to amortize index decoding and \Bpack construction, with speedups of $5.50\times$--$12.33\times$ over cuSPARSE. The rail4284 matrix has only 4.3K rows but more than 1.0M columns, with an average of 2,634 nonzeros per row. \AscSpMM achieves a $43.07\times$ speedup over cuSPARSE on this matrix. In contrast, web-Stanford, web-Google, and Delor338K contain only 8.2, 5.6, and 12.4 nonzeros per row, respectively, and achieve speedups of $1.43\times$, $1.31\times$, and $1.93\times$. Their short reductions provide less Cube computation per task. These results indicate that performance depends on the effective work per row and the locality of accesses to rows of $\matB$, rather than global sparsity alone.

\begin{figure*}[!t]
\centering
\includegraphics[pagebox=cropbox,width=0.98\textwidth]{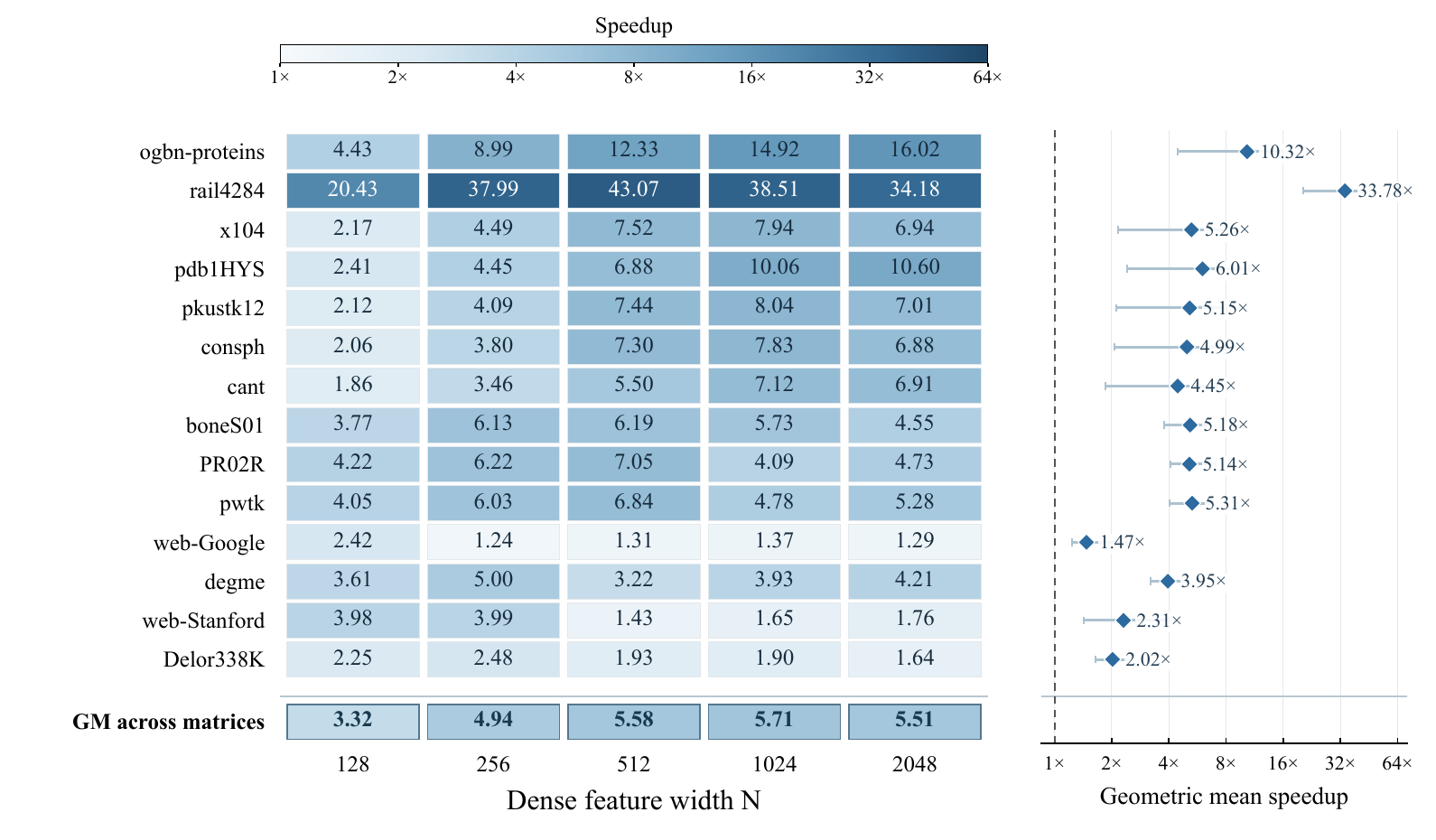}
\Description{AscSpMM speedups over A100 cuSPARSE for 14 unstructured matrices and five dense feature widths, with geometric-mean summaries.}
\caption{\AscSpMM speedups over cuSPARSE across feature widths.}
\label{fig:feature-width}
\end{figure*}

To further examine the effect of feature width, Fig.~\ref{fig:feature-width} shows the speedups of \AscSpMM over cuSPARSE for different values of $N$. We vary $N$ independently while keeping each matrix's row and column dimensions and sparsity pattern unchanged. \AscSpMM outperforms cuSPARSE on the A100 in all 70 configurations, with a geometric-mean speedup of $4.92\times$ and a range of $1.24\times$--$43.07\times$. At $N=128$, \AscSpMM remains faster on all 14 matrices, with a geometric-mean speedup of $3.32\times$. Smaller feature widths offer fewer multiply--accumulate operations per nonzero and less reuse of sparse indices and values, making sparse-processing overhead harder to amortize. Larger $N$ provides more opportunities for contiguous data movement and operand reuse, but the geometric-mean speedup rises to $5.71\times$ at $N=1024$ and then decreases to $5.51\times$ at $N=2048$. The advantage therefore depends jointly on feature width, sparsity pattern, and the balance between operand preparation and matrix computation.

\begin{figure}[!t]
\centering
\includegraphics[pagebox=cropbox,width=\columnwidth]{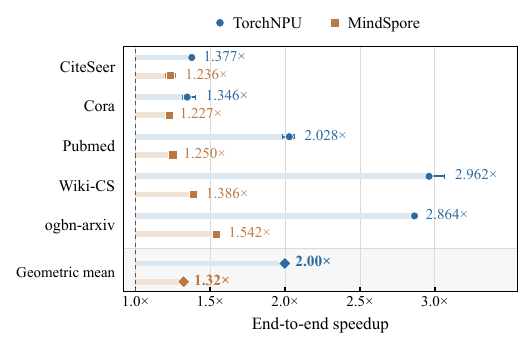}
\Description{End-to-end GCN inference speedups on five graph datasets, relative to the respective TorchNPU and MindSpore baselines, with geometric means of 2.00 and 1.32 times.}
\caption{End-to-end GCN inference speedups within each framework.}
\label{fig:gcn-end-to-end}
\end{figure}

\textbf{End-to-end performance.} To evaluate the application-level benefits of \AscSpMM, we compare GCN inference within TorchNPU and MindSpore on the same Ascend NPU. The TorchNPU baseline performs weighted neighbor aggregation through gather, elementwise multiplication, and \texttt{scatter\_add\_}, reusing indices, edge weights, and output buffers. The MindSpore baseline uses GraphLearning's COO weighted neighbor sum, compiled at O2. Within each framework, we keep the input graph, model weights, and remaining operators unchanged and replace only the two graph aggregations with \AscSpMM. We measure complete forward-pass latency, including required input padding/copying, output clearing, layout conversion, and synchronization.

As shown in Fig.~\ref{fig:gcn-end-to-end}, \AscSpMM accelerates complete inference on all five datasets in both frameworks. The geometric-mean speedups are $2.00\times$ in TorchNPU and $1.32\times$ in MindSpore, with ranges of $1.35\times$--$2.96\times$ and $1.23\times$--$1.54\times$, respectively. Prediction agreement is 100\% in both frameworks. Test accuracies remain unchanged at 70.30\%, 81.50\%, 79.70\%, 79.46\%, and 68.17\% on CiteSeer, Cora, Pubmed, Wiki-CS, and ogbn-arxiv, respectively. These results show that \AscSpMM translates sparse-operator improvements into faster complete GCN inference while preserving prediction quality across both evaluated frameworks.

\subsection{Sparsity Sensitivity and Performance Boundaries}\label{sec:sensitivity}

\begin{figure}[!t]
\centering
\includegraphics[pagebox=cropbox,width=\columnwidth]{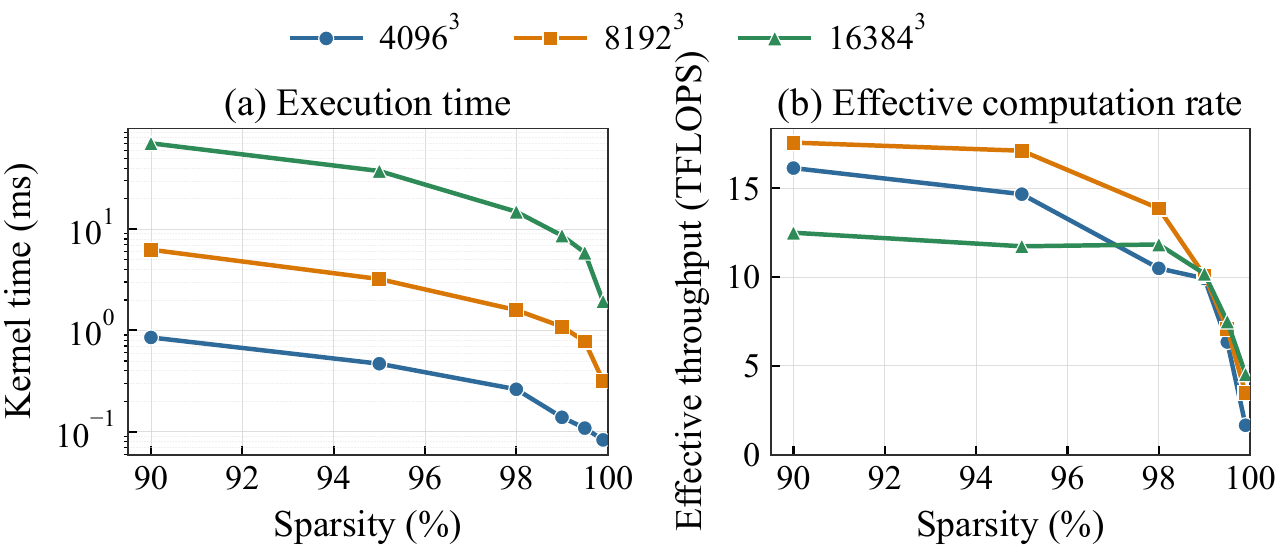}
\Description{Kernel execution time and estimated effective throughput from configured sparsity for three matrix sizes.}
\caption{Execution time and estimated effective throughput of \AscSpMM versus sparsity.}
\label{fig:sensitivity}
\end{figure}

Fig.~\ref{fig:sensitivity} shows that sparsity affects execution time and estimated effective throughput differently. As sparsity increases from 90\% to 99.9\%, execution time decreases for all three matrix sizes. Estimated effective throughput is highest at 90\% sparsity for each size and generally decreases as sparsity increases. The maximum is 17.56 TFLOPS on the $8192^3$ matrix. Throughput is not strictly monotonic: for example, it increases slightly from 95\% to 98\% sparsity on the $16384^3$ matrix. At high sparsity, latency decreases substantially, but the useful multiply--accumulate workload shrinks faster than operand preparation and data movement. Higher sparsity therefore reduces execution time, whereas the lower-sparsity configurations generally sustain higher estimated effective throughput.

\subsection{Memory-Footprint Analysis}\label{sec:memory}

\begin{figure}[!t]
\centering
\includegraphics[pagebox=cropbox,width=\columnwidth]{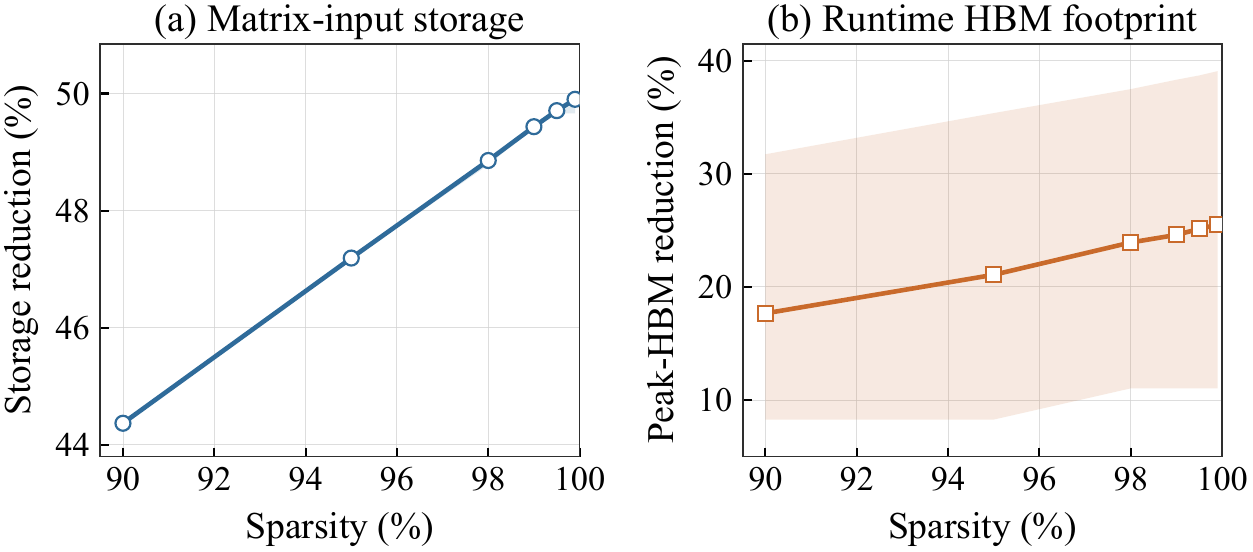}
\Description{Matrix-input storage and runtime HBM-footprint reductions versus sparsity.}
\caption{Reductions in (a) matrix-input storage and (b) peak HBM footprint; lines show means and shading shows the minimum--maximum range.}
\label{fig:memory}
\end{figure}

Figs.~\ref{fig:memory}(a) and (b) show the reductions in matrix-input storage and peak NPU HBM footprint achieved by \AscSpMM relative to AscGEMM, respectively. These memory benefits arise from both the sparse representation and the execution workspace. At the representation level, Vector-CSR avoids storing full values for zero entries in $\matA$ and allows the $m$ positions associated with the same active column within a row block to share a single column index, thereby reducing both value and metadata storage. Across the 48 configurations, \AscSpMM reduces matrix-input storage by an average of 48.24\%. The average reduction increases from 44.37\% at 90\% sparsity to 49.90\% at 99.9\%, indicating that the index and vector-block padding overheads of Vector-CSR remain well controlled across the evaluated workloads.

The reduction in peak runtime HBM footprint is smaller than that in input storage because the device must additionally accommodate the output matrix $\matC$, the \Bpack workspace, and fixed buffers. \AscSpMM packs only the $\matB$ data required by the current row block and $N$-dimensional tile, and cyclically reuses this workspace through the double-slot ring buffer. A slot can be returned to AIV immediately after AIC completes its final GM read, allowing the required workspace capacity to be determined by the pipeline depth rather than by the size of a complete intermediate matrix. After accounting for these additional allocations, \AscSpMM reduces peak HBM footprint by an average of 22.98\% across the 48 configurations. The average reduction increases from 17.66\% at 90\% sparsity to 25.50\% at 99.9\%. For the $16384^3$ matrix at 99.9\% sparsity, the reduction reaches 39.08\%.

The value of this memory optimization extends beyond reducing the storage cost of an individual operator invocation. A lower peak HBM footprint enlarges the range of SpMM problems that can be executed, allowing the NPU to process larger $M$, $K$, or $N$ dimensions under the same HBM capacity without encountering out-of-memory (OOM) failures caused by the simultaneous residency of input matrices and intermediate buffers. In end-to-end graph analytics, sparse neural networks, and scientific-computing applications, the released capacity can also be used to store node features, model parameters, and other intermediate states, or to support larger batches and greater task concurrency. Consequently, even when some moderately sparse configurations do not outperform GEMM in execution time, \AscSpMM can still expand the feasible problem scale through its memory-capacity advantage.

\subsection{Numerical Correctness}\label{sec:correctness}

\begin{figure}[!t]
\centering
\includegraphics[pagebox=cropbox,width=\columnwidth]{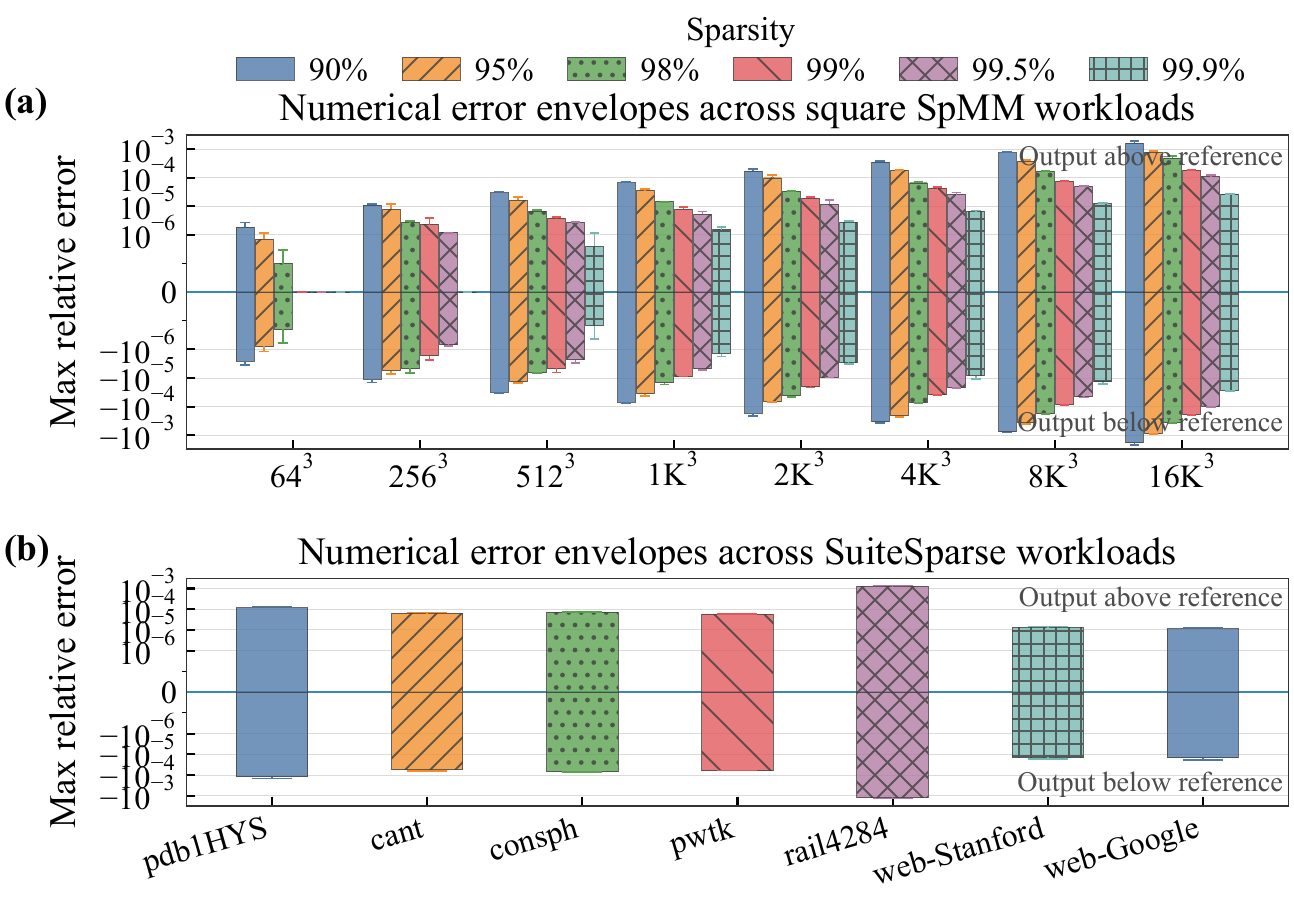}
\Description{Signed relative-error envelopes for controlled and SuiteSparse workloads.}
\caption{Signed relative-error envelopes against CPU SpMM; bars show mean error bounds, and error bars show the extrema.}
\label{fig:correctness}
\end{figure}

We validate numerical correctness by comparing the output of \AscSpMM element by element against the result of an equivalent CPU-based SpMM implementation. Fig. ~\ref{fig:correctness} reports the relative errors of \AscSpMM across the 48 controlled matrices and seven SuiteSparse matrices.

Across all tests, the largest positive and negative relative errors are $1.919\times10^{-3}$ and $-2.147\times10^{-3}$, respectively. Both occur for the $16384^3$ at 90\% sparsity, which has the largest number of accumulated terms. The largest mean relative error is $2.172\times10^{-6}$, observed on the highly irregular rail4284, while the mean relative error averaged across all tests is $2.931\times10^{-7}$. The error range gradually increases with matrix size and the number of accumulated terms along the $K$ dimension, but the positive and negative errors remain of comparable magnitude. Increasing sparsity shortens the reduction length and correspondingly narrows the error range. This distribution is consistent with changes in the ordering of FP16 multiply--accumulate operations and the accumulation of rounding errors. No persistent positive or negative bias is observed, indicating that the optimized asynchronous execution and buffer reuse preserve numerical correctness.

\subsection{Discussion}\label{sec:discussion}

The experimental results show that \AscSpMM performance is not determined by sparsity alone. Instead, it depends on the balance between the amount of effective computation contained in each matrix-computation task and the data-preparation cost required to construct that task. When active columns exhibit high occupancy within a row block, reductions along the $K$ dimension are sufficiently long, the $\matA$ operand can be reused across the $N$ dimension, and accesses to rows of $\matB$ have favorable locality, the costs of index decoding and \Bpack construction can be effectively amortized over consecutive Cube operations. In contrast, extremely short sparse rows, scattered accesses to $\matB$, or numerous underfilled tail tiles reduce vector-block occupancy and the amount of Cube computation per task, causing synchronization and data movement to become dominant again.

The central role of \AscSpMM is therefore not merely to compress sparse matrices, but to transform index-driven data selection into regularly shaped operands that Cube can consume continuously, while matching the production rate of AIV with the consumption rate of AIC through pipelining. These design principles can be extended to other NPUs with similar execution organizations \cite{liao2021ascend,jouppi2017tpu}.

\section{Conclusion}\label{sec:conclusion}

We present \AscSpMM, an efficient SpMM design for Ascend NPUs that decouples index-driven irregular data preparation from high-throughput matrix multiply--accumulate operations through a parameterized Vector-CSR representation, Vector--Cube cooperative computation, and asynchronous cross-engine pipelining. Our results demonstrate that NPUs are not limited to regular dense computation. By jointly designing the sparse representation, task mapping, and operand-delivery path, their high-throughput matrix capabilities can be extended to irregular sparse workloads. \AscSpMM demonstrates a viable path to extending NPU sparse-computation capabilities through cross-engine cooperation. Its principle of using the vector engine to resolve irregularity and the matrix engine to perform the primary computation can inform the development of high-performance sparse operators on accelerators with similarly decoupled execution paths.

\section*{Acknowledgments}
This work was partially supported by Xian Jiaotong University Kunpeng \& Ascend Center of Cultivation. Generative AI was used only to assist with English editing; all technical content and results were independently verified by the authors.

\IEEEtriggeratref{0}
\IEEEtriggercmd{\newpage}
\bibliographystyle{IEEEtran}
\bibliography{references}
\end{document}